\documentclass{elsart}
\usepackage{graphicx}
\usepackage{dcolumn}
\usepackage{bm}
\usepackage{amsmath}
\usepackage{comment}

\usepackage{color} 

\newcommand{\vect}[1]
{
\mathbf{#1}
}

\newcommand{\chem}[1]
{
\ensuremath{\mathrm{#1}}
}

\begin{document}
\begin{frontmatter}

\title{A new way of producing electron vortex probes for STEM}
\author[emat]{J. Verbeeck}
\ead{jo.verbeeck@ua.ac.be}
\author[emat]{H. Tian}
\author[emat,fei]{A. B\'ech\'e}
\address[emat]{EMAT, Electron Microscopy for Materials Science, University of Antwerp, Groenenborgerlaan 171, 2020 Antwerp, Belgium}
\address[fei]{FEI Electron Optics, NL-5600 KA Eindhoven, Netherlands}

\begin{abstract}
A spiral holographic aperture is used in the condensor plane of a scanning transmission electron microscope to produce a focussed electron vortex probe carrying a topological charge of either $-1,0$ or $+1$. The spiral aperture design has a major advantage over the previously used forked aperture in that the three beams with topological charge $m=-1,0,1$ are not side by side in the specimen plane, but rather on top of each other, focussed at different height. This allows us to have only one selected beam in focus on the sample while the others contribute only to a background signal. In this paper we describe the working principle as well as first experimental results demonstrating atomic resolution HAADF STEM images obtained with electron vortex probes. These results pave the way for atomic resolution magnetic information when combined with electron energy loss spectroscopy.
\end{abstract}


\end{frontmatter}

\section{Introduction}
Vortex waves or waves containing a phase singularity along their propagation direction were first discovered and theoretically described in the radio wave regime by Nye et al.\cite{Nye1974}. In optics, vortex beams are around since the nineties \cite{Bazhenov1991,Allen1992,Heckenberg1992,Beijersbergen,He1995} and were typically produced with a laser and some way of transforming commonly available laser modes in so-called Laguerre-Gaussian beams. Such optical vortex beams carry orbital angular momentum (OAM) of $m\hbar$ per photon around their propagation axis with $m$ the so-called topological charge \cite{Arnold2010}. This OAM has to be contrasted with the better known spin angular momentum that is carried by the polarisation of the photon wave which can take values $\pm \hbar$ \cite{Beth1935}.
Since this discovery, a wide field of applications opened up and keeps expanding. Optical vortices are used in optical tweezer setups to manipulate and rotate micron sized particles \cite{ONeil2002,Padgett2011}. As such they could be used to drive micromotors or to mix liquids on a micrometer scale. The fact that the vortex state of photons can carry in principle an infinite amount of integer values for $m$ is interesting in telecommunication were a single wavepacket of light could now carry much more information encoded in the topological charge \cite{Celechovsky2007}.
Another exciting application is the use of a so-called vortex coronograph to detect faint signals nearby bright stellar objects \cite{Swartzlander2005,Berkhout2009} and such a device was used to detect the weak reflection of exoplanets located very close to a bright star\cite{Serabyn2010}. Besides technological applications, optical vortex beams also provoke many interesting fundamental issues and more complicated arrangements of the phase discontinuity in three dimensions can be studied\cite{Dennis2010,Terriza2001}.
Apart from optics, vortex beams have been produced and used with radiowaves\cite{Thide2007}, acoustical waves\cite{Skeldon2008} and X-rays\cite{Peele2002,Cojoc2006} each having its own potential for exciting applications. 
Recently also electron vortex beams were theoretically described \cite{Bliokh2007} and experimentally observed \cite{Uchida,Verbeecknature}. We demonstrated a practical way to make such beams in a conventional TEM and to use them to obtain magnetic information in electron energy loss spectroscopy in a technique that is similar to XMCD and EMCD \cite{Verbeecknature,Schutz1987,PSnature}. As electrons are charged particles, electron vortices have similar properties as their optical counterparts but on top of the OAM of $m\hbar$ per electron, they also carry a magnetic moment of $m\mu_B$ per electron\cite{Bliokh2007}. We recently described some of the properties of electron vortex beams in an electron microscope \cite{PSvortextheory} and others have digged deeper into the theoretical interplay of OAM and spin for electrons \cite{Wang2011} and in general to the concept of OAM in particle physics\cite{Ivanov2011}.
The major attraction of electron vortex beams is the fact that they can be made of atomic size \cite{verbeeckangstrom,schattschneiderangstrom} and that they can deliver magnetic information in EELS experiments \cite{Verbeecknature}. 

A drawback in the original setup using a forked aperture  is the fact that this produces three beams with different topological charge $-m,0,+m$ \cite{Verbeecknature}. Using such a beam in a STEM setup would lead to all beams interacting simulatenously with the material unless the material is a nanoparticle in which case we can use its shape to select a specifc beam with well defined $m$\cite{Tian2011}. In the present paper we present an alternative method that solves this issue by using spiral shaped apertures. This technique was used in optics from the very start\cite{Heckenberg1992} but we will demonstrate here that it is particularly useful for electron vortices as well. We designed an appropriate spiral phase plate for TEM and demonstrate that we are able to obtain atomic resolution HAADF STEM images with a vortex beam of $m=-1,0,+1$. Although we focus here on producing $|m|=1$, it is trivial to design an aperture for any other value of $m$ whith the detailed description given here.

\section{Spiral apertures}
\begin{figure}
\includegraphics[width=0.7\textwidth]{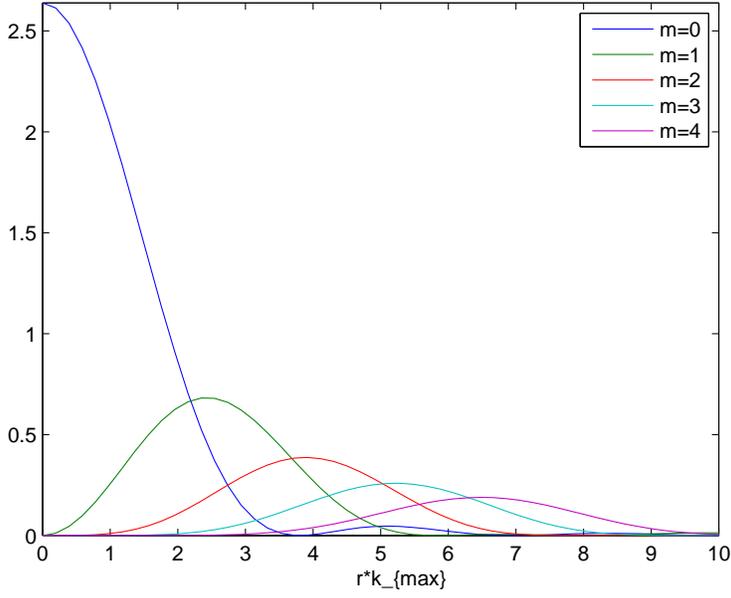}
\caption{\label{airy} Radial intensity profiles for vortex beams obtained with a flat filling of a circular aperture in the condensor plane as in eq.\ref{flatfill}. For $m=0$, the typical Airy disc profile is obtained while for $m\neq 0$ a doughnut like shape with increasing width as $m$ increases is obtained.}
\end{figure}

One way of producing electron vortex beams is using the so-called holographic reconstruction \cite{Verbeecknature}. Suppose we want a vortex beam in the condensor system of a TEM of the type:
\begin{equation}
\psi_m(k_{\perp},\phi,z)=e^{im\phi}\Pi(\frac{k_\perp}{2k_{max}})e^{ikz}
\label{flatfill}
\end{equation}
With $k_{max}=k_0\alpha$ the size of the angle limitting aperture with $\alpha$ the convergence semi-angle and $k_0$ the electron wave vector. We use $(k_\perp,\phi,z)$ as cylinder coordinates in the condensor plane and $\Pi(x)$ the rectangle function which equals 1 for $|x|<\frac{1}{2}$.
In the specimen plane this becomes after Hankel transforming and neglecting aberrations in the condensor system (for a discussion on the effect of aberrations on vortex beams see \cite{schattschneiderangstrom}):
\begin{equation}
\psi_m(r,\phi,z)=e^{im\phi} e^{ikz} \int_0^{k_0\alpha} J_m(k_\perp r) k_\perp dk_\perp  
\end{equation}
For $m=0$ this leads to the well known Airy disc function. For higher $|m|$, a doughnut-like profile is obtained as displayed in fig.\ref{airy}.

In holographic reconstruction we reconstruct this wave by obtaining an aperture that is calculated as an interference pattern between the desired wave and a reference wave. A commonly used reference wave is a tilted plane wave:
\begin{equation}
\psi'_{ref}(\vect{k_\perp},z)=e^{i\vect{k_\perp}.\vect{d}}e^{ikz}
\end{equation}
With $\vect{d}$ expressing the tilt. Using such a tilted plane wave leads to a typical forked interference pattern\cite{Verbeecknature,Bazhenov1991}. Illuminating such a forked pattern leads to the creation of a set of three beams displaced by $\vect{d}$ with respect to each other in the far field: the wanted beam $\psi_m(\vect{r}-\vect{d})$, the reference wave $\psi_{ref}(\vect{r})$ wave and the complex conjugate of the wanted beam $\psi_{-m}(\vect{r}+\vect{d})$. Having all three beams present at the same time in the specimen plane is undesirable because it makes scanning approaches impossible (image resolution would suffer) and it does not allow to obtain EELS spectra stemming from only one of the probes. One could think of designing apertures that can block the unwanted parts of the probe, but for the given configuration of the condensor system in a modern TEM, this is far from trivial.

A solution to this problem is to use a different reference wave \footnote{any reference wave with a phase varying with $k_\perp$ is in principle possible}:
\begin{equation}
\psi_{ref}(k_\perp,\phi,z)=e^{i\frac{\pi}{\lambda} (\frac{k_\perp}{k_0})^2  \Delta f}e^{ikz}\Pi(\frac{k_\perp}{2k_{max}})
\end{equation}

\begin{figure}
\includegraphics[width=0.5\textwidth]{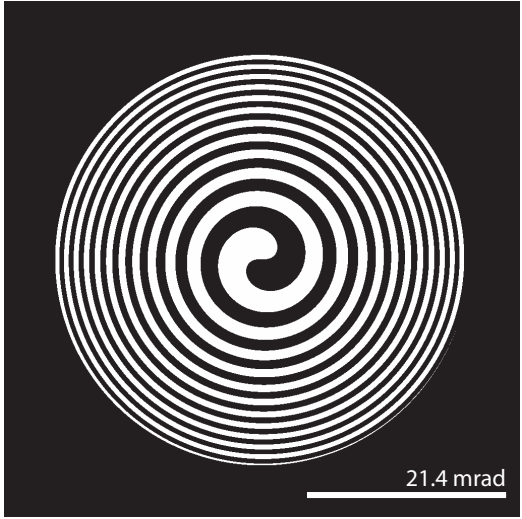}
\caption{\label{spiral} Spiral aperture design for $m=1$, $\Delta f=100$~nm, $E_0=300$~kV and $\alpha=21.4$~mrad.}
\end{figure}
\begin{figure}
\includegraphics[width=0.5\textwidth]{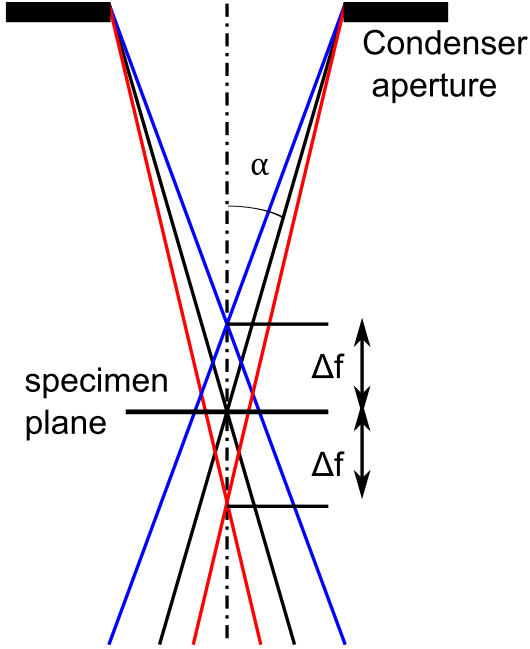}
\caption{\label{sketch} Sketch of the three beams with different topological charge each with a different focal plane.}
\end{figure}

This is a typical thin round lens function with a focal length of $\Delta f$. Interfering this with $\psi_m$ leads to a spiralling interference pattern as depicted in fig.\ref{spiral} and is given by:
\begin{eqnarray}
I_{fringe}(k_\perp,\phi)&=&|\psi_m+\psi_{ref}|^2\nonumber\\
&=&|\psi_m|^2+2Re(\psi_m\psi_{ref}^*)+|\psi_{ref}|^2\nonumber\\
&=&2\Pi(\frac{k_\perp}{2k_0\alpha})+2\cos(m\phi+\Delta f \pi(\frac{k_\perp}{k_0})^2)
\label{spiraleq}
\end{eqnarray}
As it is difficult to make a continuous amplitude aperture for TEM, we discretize this fringe pattern to become either transparent for $I_{fringe}>2$ or opaque for $I_{fringe}\leq 2$. This approximation will lead to some higher order effects, but for the rest of the derivation we will neglect these.
Illuminating this fringe pattern with a plane reference wave and transforming to the specimen plane we get:
\begin{eqnarray}
\psi_{spec}(r,\phi)&=&\int_0^{k_0\alpha} J_0(k_\perp r)k_\perp dk_\perp\nonumber \\
&+&\frac{1}{2}e^{im\phi}\int_0^{k_0\alpha} J_m(k_\perp r)e^{-i\Delta f \pi(\frac{k_\perp}{k_0})^2}k_\perp dk_\perp\nonumber\\
&+&\frac{1}{2}e^{-im\phi}\int_0^{k_0\alpha} J_{-m}(k_\perp r)e^{+i\Delta f \pi(\frac{k_\perp}{k_0})^2}k_\perp dk_\perp
\label{rvortex}
\end{eqnarray}
The three terms are now again three separate probes, the first term is an Airy disc and the two other terms are two opposite vortex beams with topological charge of $m$ and $-m$ but defocussed over $-\Delta f$ and  $+\Delta f$. So, instead of having three beams next to each other, we now have three beams on top of each other separated by a defocus distance $\Delta f$ that can be chosen while designing the aperture. A sketch of these three beams is given in fig.\ref{sketch}.

\begin{figure}
\includegraphics[width=0.7\textwidth]{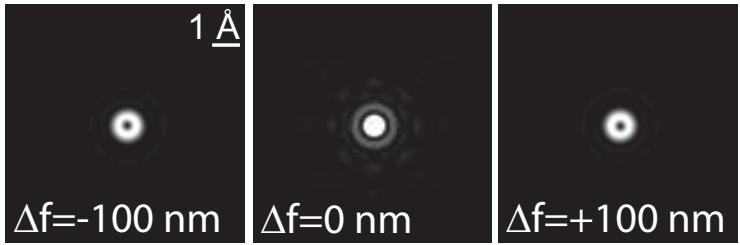}
\caption{\label{simulated}Simulated intensity in the specimen plane for a defocus of $\Delta f=-100,0,+100$~nm for 300~kV and $\alpha=21.4$~mrad and m=1. The aperture used in this simulation contains the reinforcement bars as in fig.\ref{fib} showing that their effect is not detrimental.}
\end{figure}

The simulated pattern obtained when illuminating the spiral pattern is shown in fig.\ref{simulated} for three different defocus values, a through focal series movie is available as supplementary information\cite{supp}. This nicely shows that by defocussing the condensor system one can choose between having either the -m,0 or +m vortex beam in focus. Note that in all z-planes, the total angular momentum of the electron wave remains zero, but the spatial localisation of the three constituent waves is different. Using such a wave in a STEM-EELS setup would lead to spatial information only coming from the wave which is in focus while the other two parts will contribute to a background signal. In fact, one can estimate the signal to background ratio e.g. in a situation with one point scatterer per square unit cell of size a by a (an infinitely thin sample). The signal intensity $I_m$ coming from the probe in focus located on a point scatterer can be approximated as the inverse of the area of that probe times the 1/2 weight:
\begin{equation}
I_m\approx \frac{(k_0\alpha)^2}{2\pi}
\end{equation}
The background signal consists of the probe with topological charge $0$ and $-m$ with a weight of respectively $1$ and $1/2$. We assume that this signal is evenly spread over the unit cell (if $\Delta f \alpha >> a$) and we get a signal height $I_b$ of:
\begin{equation}
I_b\approx \frac{3}{2a^2}
\end{equation}
The amplitude of the spatially varying signal on this uniform background will then be a measure of the signal to background ratio:
\begin{equation}
\frac{I_m}{I_b}=\frac{(k_0\alpha a)^2}{3\pi}
\end{equation}
Using this approximation one can see that for a unit cell parameter of $a=4$\AA~ and a typical convergence angle of $\alpha=20$~mrad a signal to background ratio of $67$ is obtained. The above approximation is a worst-case scenario where all targeted atoms of a certain type are in the same magnetic state (leading to the same OAM dependent EELS signal) and does not explicitly include delocalisation effects. In fact, a detailed description of inelastic scattering of vortex beams is far beyond the scope of this paper and will be treated elsewhere \cite{verbeecktheory}. Note that this method relies on the presence of spatial variation in the EELS signal and therefore requires crystalline samples with projected interatomic distances between specific atoms that can be resolved in (atomic resolution) EELS maps.

Although $\Delta f$ does not appear in the above approximation, it plays an important role as we want the sample to have a thickness that is considerably thinner than this defocus to avoid mixing the signals. Note that for a given spiral aperture design, the defocus depends on the convergence angle that is chosen by setting the strength of the condensor lenses and scales as $1/\alpha^2$. Spiral apertures with a high defocus value and a high convergence angle become difficult to produce since they contain finer and finer features which put strong requirements on the performance of the FIB instrument.

\section{Experimental results}
\begin{figure}
\includegraphics[width=0.5\textwidth]{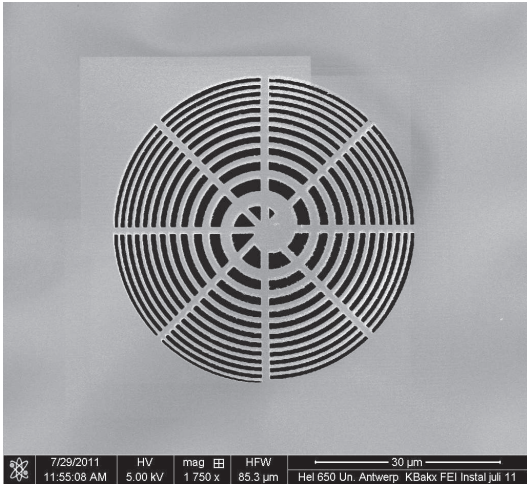}
\caption{\label{fib} SEM image of the $50 \mu m$ spiral aperture made by FIB including mechanical reinforcement bars.}
\end{figure}

A spiral aperture according to fig.\ref{spiral} is produced by sputtering a thin layer of a few hundred nm of \chem{Pt} on a 50~nm commercially available \chem{Si_3N_4} window and using a dual beam FIB to cut out the pattern to a diameter of $50~\mu m$. Extra reinforcement bars are added for mechanical stability and the result is shown in the SEM image in fig.\ref{fib}. The reinforcement bars were taken into account in the simulation of fig.\ref{simulated} and have surprisingly little effect on the produced pattern. 

\begin{figure}
\includegraphics[width=0.7\textwidth]{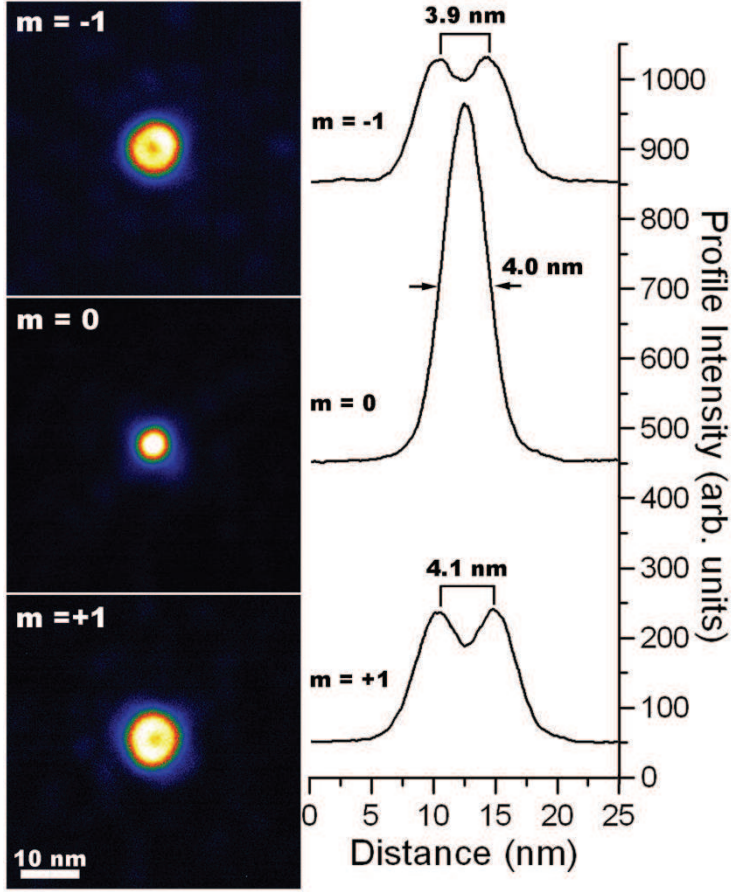}
\caption{\label{expvortex} Image of the vortex beam for three different defoci of $-33,0,+37 \mu m$ at a convergence semi-angle of approximately $\alpha=0.4$~mrad and at 300~kV. Line profiles in the right panel show the doughnut like profiles for $m=\pm 1$ and the Airy disc profile for $m=0$. The aperture is simmilar to the one presented in fig.\ref{fib}.}
\end{figure}

Introducing this in the condensor aperture of a Philips CM30 microscope operating at 300 kV leads to real space intensity images at different defoci shown in fig.\ref{expvortex}. In order to avoid issues with lens aberrations, a very low convergence of approximately $\alpha=0.4$~mrad was chosen. This increases the required experimental defocus to a nominal value of approximately $\Delta f=\pm 30\mu m$. The intensity profiles show the typical dougnut like pattern with a diameter of approximately 4~nm. Note that the intensity in the center of the doughnut does not approach zero which is due to the partial coherence of the source as explained in detail in \cite{Swartzlander2007} for photons and in \cite{verbeeckangstrom,schattschneiderangstrom} for electrons. This effect leads to a probe with a mixed OAM state in the center of the doughnut while further away from the center the pure OAM state remains\cite{Swartzlander2007}. Adjusting the demagnification of the source via the spot size control of the microscope allows to increase the partial coherence at the expense of probe current.

\begin{figure}
\includegraphics[width=1.0\textwidth]{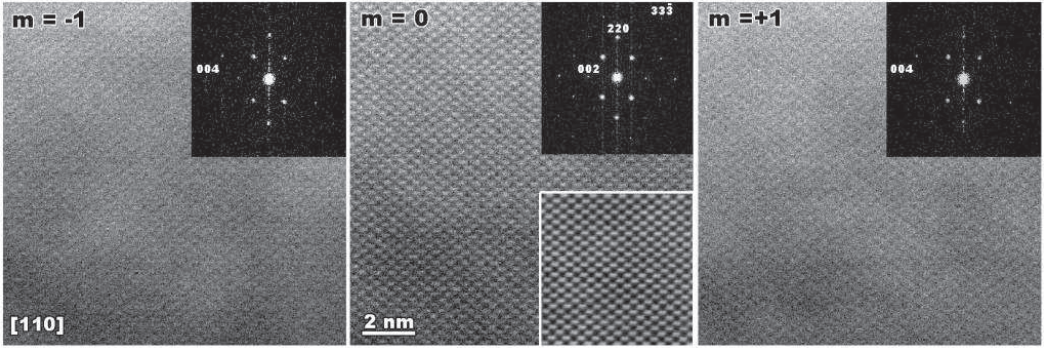}
\caption{\label{stemvortex} HAADF STEM images of [110] \chem{GaAs} obtained with either the $m=-1,0,+1$ in focus at $\Delta f=-125,0,+114$~nm showing the ability to obtain high resolution STEM images with probes containing topological charge. Insets are diffractograms which allow to estimate the attainable resolution. The lower inset in the $m=0$ image is an HAADF STEM image obtained with a conventional round aperture of $50 \mu m$, it shows almost identical resolution but much better signal to noise ratio. Microscope parameters are $\alpha=15.5$~mrad and 300~kV.}
\end{figure}

The above experiment proves that the concept works, but in order to reach higher spatial resolution and STEM capabilities we inserted the spiral aperture into the C2 aperture of the double corrected FEI Titan$^3$ Qu-Ant-EM microscope operating at 300~kV. Doing so, enables us to increase the convergence angle to $\alpha=15.5$~mrad and to obtain HAADF STEM images with a detector inner angle of approximately 100~mrad. Fig.\ref{stemvortex} shows that we still obtain atomic resolution images on a [110] \chem{GaAs} sample for three distinct focus settings of $\Delta f=-125,0,114$~nm. Note that the atomic lattice image dissapears completely before reappearing when defocussing from say the $m=-1$ to the $m=0$ focussed plane, in agreement with our simulations. Diffractograms in fig.\ref{stemvortex} show the best resolution for the $m=0$ probe up to 1.088 \AA ($33\bar{3}$) and up to 1.413 \AA (004) for the $m=\pm 1$ probe. The fact that the obtained resolution for the $m=\pm 1$ probe is lower than the m=0 beam can be understood when looking back at fig.\ref{airy}. For higher $m$ values this effect is expected to be more pronounced.
These experiments demonstrate the ability to perform atomic resolution STEM experiments with vortex beams. Further resolution enhancements are expected when increasing the convergence semi-angle but this mainly depends on how well the aberrations of the probe forming lenses can be corrected. Imaging the probe as in fig.\ref{expvortex} requires on top of this an excellent allignment of the image corrector. 

\section{Discussion}
The design and use of a spiral aperture was demonstrated at both low and high convergence angle settings and qualitative agreement with simulations is obtained. Great care needs to be taken in making the aperture as clean as possible with no remaining debris from the FIB cutting as this debris will act to randomise the phases in the condensor plane. Care is also needed to limit the virtual source size of the microscope by choosing a small spot setting to obtain a coherent illumination of the condensor aperture. We show that all these effects can be kept under control and enough beam current is available to do atomic resolution HAADF STEM. This is an important step in the direction of obtaining EELS from these focussed probes but this is another major undertaking with signal to noise ratio being of crucial importance. In that respect it is interesting to note that the aperture itself will block 50\% of all electrons and the remaining current is divided over the three probes according to eq.\ref{rvortex} leading to a total theoretical current reduction of $\frac{1}{8}$ in the beams carrying the topological charge\footnote{This is an overestimate since higher order effects from the discretised aperture will reduce this intensity further.}.

\section{Conclusion}
We have demonstrated that spiral apertures form an interesting alternative to forked apertures for making electron vortex beams. The main advantage is that for the spiral apertures, the different beams with topoligical charge $-m,0,+m$ are in focus in different planes. This allows us to use them in a scanning transmission electron microscope to obtain the first atomically resolved HAADF STEM images created with vortex electron beams. The experimental behaviour of these beams agrees well with simulations and opens the door for further atomic resolution vortex beam experiments. This is especially promising in cases where atomically resolved EELS signals can be obtained as atomic resolution magnetic information could become a reality.

\section{acknowledgments}
J.V. wants to thank Miles Padgett for suggesting this setup and pointing me to the relevant optics literature. Peter Schattschneider is acknowledged for in depth discussions on related topics. J. V acknowledges funding from the European Research Council under the 7th Framework Program (FP7), ERC grant No. 46791-COUNTATOMS and ERC Starting Grant No. 278510 VORTEX. The Qu-Ant-EM microscope is partially funded by the Hercules fund of the Flemish Governement.

\end{document}